\documentclass[conference]{IEEEtran}

\makeatletter

\def\ps@headings{%

\def\@evenhead{\scriptsize\thepage \hfil \leftmark\mbox{}}%

\def\@oddfoot{}%

\def\@evenfoot{}}

\makeatother

\usepackage{algorithmic}

\usepackage{algorithm}
\usepackage{verbatim}
\usepackage{amsmath}
\usepackage{enumerate}
\usepackage{fancybox}

\usepackage{verbatim}
\usepackage{array}
\usepackage{graphicx}
\usepackage{float}
\usepackage[footnotesize]{subfigure}
{\begin{Sbox}\begin{minipage}}%
{\end{minipage}\end{Sbox}\fbox{\TheSbox}}

\newtheorem{theorem}{Theorem}

\IEEEoverridecommandlockouts
\IEEEpubid{%
\makebox[\columnwidth]{ISBN 978-3-901882-83-8~\copyright~2016 IFIP \hfill}%
	\hspace{\columnsep}%
	\makebox[\columnwidth]{ }%
}

\begin{document}

\title{Content-Centric Networking \\
Using Anonymous Datagrams }

\author{J.J. Garcia-Luna-Aceves$^{1,2}$
and Maziar Mirzazad Barijough$^2$ \\
$^1$Palo Alto Research Center, Palo Alto, CA 94304 \\
$^2$Department of Computer Engineering,
 University of California, Santa Cruz, CA 95064\\
 Email: jj@soe.ucsc.edu, maziar@soe.ucsc.edu }

\maketitle

\begin{abstract}

Using Interests (requests that elicit content) and  maintaining 
per-Interest forwarding state in Pending Interest Tables (PIT)  are 
integral to the design of  the Named Data Networking (NDN) and Content-Centric Networking (CCNx) architectures. However,  
using PITs makes the network vulnerable to Interest-flooding attacks, and
PITs can become very large. 
It is shown that  in-network caching eliminates the need for Interest aggregation and obviates the use of PITs.
A new approach to content-centric networking (CCN-GRAM) is introduced that provides all the benefits of 
NDN and CCNx,  eliminates the use of PITs by means of anonymous datagrams, and is immune to Interest-flooding attacks. 
Routers  
maintain routes to the anonymous origins of Interests using an on-demand routing approach in the data plane that can also be used to provide native support for  multicasting in the dat a plane.   
Simulation experiments  show that the number of forwarding entries required in CCN-GRAM  is orders of magnitude smaller than the number of  PIT entries.

\end{abstract}


\section{Introduction}
 
The leading approach in content-centric networking 
consists of: populating forwarding information bases (FIB) maintained by routers with routes to name prefixes denoting content, sending content requests (called Interests) for specific content objects (CO) over paths implied by the FIBs, and delivering data packets with content objects along the reverse paths traversed by Interests. 

The main advantages that such Interest-based content-centric networking  approach offers compared to the IP Internet  are that:  (a) content providers and caching sites do not know the identity of the consumers requesting content; (b) content can be obtained by name  from 
those sites that are closer to consumers; (c) data packets carrying content cannot traverse loops, because they are sent over the reverse paths traversed by Interests; 
and (d) content-oriented security mechanisms can be implemented as part of the content delivery mechanisms.

Named data networking (NDN) \cite{ndn} and CCNx \cite{ccnx} are the two prominent   Interest-based content-centric networking  approaches.
Routers in NDN and CCNx  maintain a  ``stateful forwarding plane" \cite{ndn-fw2}  (i.e., per-Interest forwarding state) by means of Pending Interest Tables (PIT). The PIT of a router maintains information regarding the incoming interfaces from which Interests for a CO were received and the interfaces where the Interest for the same CO was forwarded. 

Since the inception of CCNx and NDN, PITs have been viewed as necessary in order to maintain routes to the origins of Interests while preserving the anonymity of those sources, aggregate Interests requesting the same content in order to attain efficient Interest and content forwarding, and support  multicasting without additional support in the control plane. 

However, using PITs at  Internet scale comes at a big price.  PITs grow very large   \cite{dai-12, tsi14, var-13} as the number of Interests from users increases, which results from PITs having to store per-Interest forwarding state. Furthermore,  PITs 
make routers vulnerable to  Interest-flooding attacks \cite{DDos1, vir-13, wahl13a, wahl13b} in which adversaries send malicious Interests aimed at  making the size of PITs explode.   Known  countermeasures  to these attacks \cite{afan} attempt to reduce the rates at which suspected routers can forward  Interests. However, these solutions cannot prevent all flooding attacks and can actually be used to mount other types of denial-of-service attacks. 

Section~\ref{sec-prelim}  analyzes the effectiveness of Interest aggregation in NDN by means of simulations  based on the implementation of NDN in  ndnSIM  \cite{ndnsim} without modifications. 
The results  show that the percentage of Interests that are aggregated  is negligible when in-network caching is enabled,
even when Interests exhibit  temporal or spatial correlation.

Given that  in-network caching obviates the need for Interest aggregation, and
given the vulnerability of NDN and CCNx to Interest-flooding attacks, it is clear that a new  Interest forwarding approach is needed for content-centric networking.

We present {\bf CCN-GRAM} ({\it Gathering of Routes for Anonymous Messengers}),  which provides all the benefits of content-centric networking,
including native support for multicasting in the data plane, and 
eliminates the need to  maintain per-Interest forwarding state by  forwarding Interests and responses  to them using  {\em anonymous} datagrams. 

Section \ref{sec-design} describes the operation of CCN-GRAM. Like NDN and CCNx, CCN-GRAM uses Interests, data packets, and 
replies  to Interests.
Similar to IP datagrams, the messages sent in CCN-GRAM specify a source and a destination. For an Interest,  the source of an Interest is an anonymous identifier with local context and the destination is the name of a content object. For data packets and replies to Interests, the source is the name of a content object and the destination is an anonymous identifier. 
A novel on-demand routing approach is  used to maintain routes to the anonymous routers that originate Interests for specific content on behalf of local content consumers.  Only  the local router serving a user knows the identity of the user;  no other router, content provider, or caching site can determine the consumer that originated an Interest, without routers  collaborating along the path traversed by the Interests to establish the provenance of the Interest.

In contrast to NDN and CCNx in which Interests may traverse forwarding loops \cite{ifip2015, ancs2015, nof2015}, 
forwarding loops  cannot occur in CCN-GRAM for either Interests or responses sent to Interests, even if the FIBs maintained by routers contain inconsistent forwarding state  involving routing-table loops. Furthermore, 
the anonymous datagram forwarding of CCN-GRAM is much simpler than the label-swapping approach we have advocated before \cite{ocean15, icnc16}. 

Forwarding of Interests and responses to them in CCN-GRAM uses  four tables: a LIGHT (Local Interests GatHered Table), a FIB, an ART (Anonymous Routing Table) and a LIST (Local Interval Set Table).  The LIGHT of a router is an index  listing content that is locally available and content that is remote and has been requested by {\em local} users. The FIB of each router  states the 
next hops to each name prefix and the distance to the name prefix reported by each next hop.
The ART is  maintained  using Interests and states the 
paths  to destinations denoted with local identifiers from which  routers cannot discern the origin  of Interests.
The LIST states the intervals of local identifiers that a router assigns to its neighbors and that each neighbor assigns to the router.


Section \ref{sec-perf}  compares the performance of CCN-GRAM with NDN when routes to name prefixes are static and loop-free, which is the best case for NDN. 
The network consists of 150 routers, with 10 being connected to content producers and 50 being connected to consumers. 
CCN-GRAM attains similar end-to-end latencies than NDN in retrieving content. However, depending on the rate at which Interests are submitted, CCN-GRAM  requires an average number of forwarding entries  per router
that is 5 to more than150 times smaller than the number of PIT entries needed in NDN.

\section{Interest Aggregation in NDN }
\label{sec-prelim}

\subsection{Elements of NDN Operation}

Routers in NDN use Interests, data packets, and negative acknowledgments (NACK)  to exchange content.
An Interest is identified in NDN by the name of the CO requested and a nonce created by the origin of the Interest. A data packet includes the CO name, a security payload, and the payload itself. A NACK carries the information needed to denote an Interest and a code stating the reason for the response. 

A router $r$ uses three data structures  to process Interests, data packets, and NACKs:
A  content store ($CS$), 
a forwarding information base ($FIB$),  and a pending Interest table ($PIT$). 
A $CS$ is a cache for COs indexed by their names. With on-path caching, routers cache the content they receive in response to Interests they  forward.

A $FIB$ is populated using content routing protocols \cite{dcr, nlsr} or static routes and  a router matches Interest names stating a specific CO   to $FIB$ entries 
corresponding to prefix names using \emph{longest prefix match}. 
The FIB entry  for  a given name prefix  lists the interfaces that can be used to reach the prefix. In NDN, 
a  FIB entry also contains additional information \cite{ndn}.


The entry in  a $PIT$ for a given CO 
consists of  one or multiple tuples stating a nonce received in an Interest for the CO,  the incoming interface where it was received, and a list of the outgoing interfaces over which the Interest was forwarded. 


When a router receives an Interest, it checks whether there is a match in its CS for the CO requested in the Interest. The Interest matching mechanisms used can vary, and for simplicity we focus on  exact Interest matching only.  If a match to the Interest is found, the router sends back a data packet over the reverse path traversed by the Interest. If no match is found in the CS, the router determines whether the PIT stores an entry for the same content.   

In NDN, 
if the Interest states a nonce that differs from those stored in the  PIT entry for the requested content, then the router ``aggregates" the Interest by adding the incoming interface from which the Interest was received and the nonce to the PIT entry without forwarding the Interest.  If the same nonce in the  Interest is already listed in the PIT entry for the requested CO, the router sends a NACK over the reverse path traversed by the Interest.
If a router does not find a match in its CS and PIT, the router forwards the Interest along a route (or routes) listed  in its FIB for the best prefix match.     In  NDN, a router can select an  interface  to forward an Interest if it is known that it can bring content  and its performance is ranked higher than other interfaces that can also bring content. 

\subsection{Likelihood of Interest Aggregation in NDN}
\label{sec-maziar}

We analyze  the likelihood that   interest aggregation occurs in the presence of in-network caching  in NDN 
using simulations carried out with the  NDN implementation in ndnSIM  \cite{ndnsim}without  modifications. 
A more detailed analysis is presented in \cite{ali-ifip16}.
Our study is  independent of the Interest retransmission strategy, and uses the percentage of aggregated Interests  in the network 
as the performance metric.  
For simplicity, we assume that routers use exact Interest matching to decide whether an Interest can be answered.

\subsubsection{Scenario Parameters}

We consider the average latencies between routers, the  capacity of  caches, the Interest request rates from routers,  the popularity of content, and the temporal correlation of content requests. 
The scenarios we use consist of random networks with 200 nodes corresponding to routers distributed uniformly in a 100m$\times$100m area. 
Routers with 12m or shorter distance are connected to each other with a point-to-point link, which results in  a topology with 1786 edges.  
Each  router  acts as  a producer of content and also has local consumers generating Interests.   

Producers are assumed  to publish 1,000,000 different content objects that are uniformly distributed among routers. 
For simplicity, we assume that {\it all} routers have the same storage capacity in their caches, which depending on the experiment ranges  from 0 to  up to 100,000 cache entries per router, or 10\% of the published objects.

The distribution of object requests  determines how many Interests from different users request the same content.                                                                                                                                                                                                                                                                                                                                                                                                                                                                                                                                                                                                                                                                                                                                                                                                                                                                                                                                                                                                                                                                                                                                                                                                                                                                                                                                                                                                                                                                                                                                                                                                                                                                                                                                                                                                                                                                                                                                                                                                                                                                                                                                                               It has been argued  \cite{fri12} that Internet traffic follows a Zipf distribution with  a   parameter ($\alpha$) value  close to 1. A smaller Zipf parameter value results in a lower Interest aggregation amount.  Accordingly,
we model object popularity using  a Zipf distribution with  values of $\alpha$ equal to  0.7 and 1. 

We considered different values of the {\it total Interest rate per router}, corresponding to the sum of Interests from all local users. Increasing values of Interest rates can be viewed as higher request rates from a constant user population of local active users per router, or an increasing  population of active users per router. For example,  50 to 500 Interests  per second per router can be just 10 Interests per second per active user for  a local population of 5 to 50 concurrently active users per router. The Interest rates we assume per router are not large  compared to recent results on the size that PITs would have in realistic  settings  \cite{dai-12, var-13, vir-13, tsi14}. 

The percentage of Interests that benefit from Interest aggregation 
is a function of the 
RTT between the router originating the Interest and the site with the requested content, as well as the PIT entry expiration time when the Interest is not answered with a data packet or a NACK. 
Recent Internet latency statistics \cite{att}  show that  Internet traffic latency varies from 11ms for regional European  traffic to 160ms for long-distance traffic. Accordingly, we consider  point-to-point delays of 10ms between neighbor routers
in many of our simulations, which leads to  RTTs of about  200ms. We also carried experiments varying the RTT of the network below and above 200ms.

\subsubsection{Simulation Results}

The following simulation results can be viewed as applicable to the steady-state behavior of a network using NDN. 

\begin{figure}[h]
\begin{centering}
    \mbox{
    \subfigure{\scalebox{.42}{\includegraphics{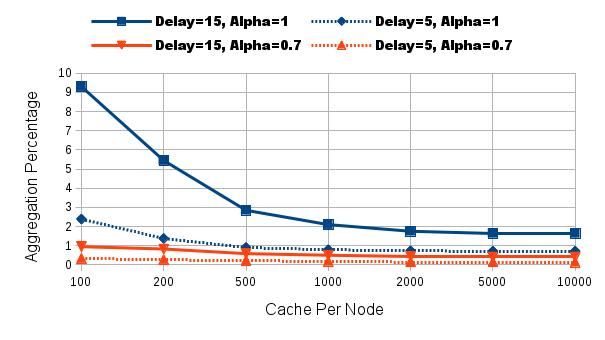}}}
      }
\vspace{-0.34in}
   \caption{Interest aggregation 
   as a function of values of Zipf parameter, caching capacity, and RTTs
   }
   \label{alpha-rtt}
\end{centering} 
\end{figure}

Figure \ref{alpha-rtt} shows the effect of the $\alpha$ parameter and RTTs when the request rate per router is 50 Interests per second. 
The latencies between neighbor routers are set to 5 and 15 ms, which produce RTTs  of 66 to 70 ms and 193 to 200 ms, respectively. 
It is clear that  Interest aggregation is far less important when consumers are less likely to request similar content ($\alpha = 0.7$).
Furthermore, the benefits of Interest aggregation vanish as caches are allowed to cache more content. When caches can store  up to 1\% of the total number of objects published, the percentage of Interests that are  aggregated  is less than 2\% for 
$\alpha = 1$ and less than  0.8\% for $\alpha = 0.7$.

\begin{figure}[h]
\begin{centering}
    \mbox{
    \subfigure{\scalebox{.37}{\includegraphics{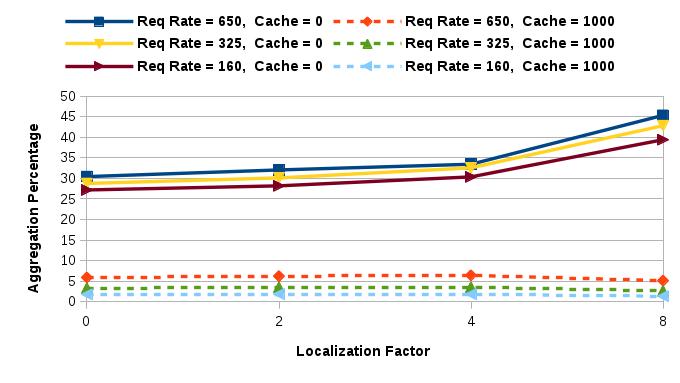}}}
      }
\vspace{-0.3in}
   \caption{Percentage of Interest aggregation under temporal locality}
   \label{localized}
\end{centering} 
\end{figure}

In theory, Interest aggregation is most useful when Interests exhibit temporal correlation, such as when popular live events take place. 
Figure \ref{localized}  shows the impact of caching on Interest aggregation when Interests have temporal correlation and either 
no caching is used or caches with capacity for only 0.1\% of the objects published in the network are used.
Interests  are generated using the model proposed by  Dabirmoghaddam et al  \cite{caching} with  a Zipf parameter value of $\alpha = 0.7$ and  results for three total Interest rates per router and four temporal localization factors for Interests are shown.
A  higher temporal locality  factor  indicates a higher degree of popularity of objects in 
the same time period. The results in Figure \ref{localized} show that, without caching, 
Interest aggregation  {\em is}  very important for all values of temporal locality of Interest popularity, and is more important when Interest locality is high (large localization factor). However, once caching is allowed and even if caches can store only up to 0.1\% of the  published objects,  the percentage of aggregated Interests is minuscule and  actually decreases with the temporal correlation of Interests. 

 \vspace{-0.06in}

\vspace{-0.24in}

\section{CCN-GRAM }
\label{sec-design}


We assume that Interests are retransmitted only by the consumers that originated them.
We assume  that routers use exact Interest matching, and that a router that advertises being an origin of a name prefix stores  all the content objects associated with that prefix at a local content store.
Routers know which interfaces are neighbor routers and which are local users, and forward Interests on a best-effort basis. For convenience, it is assumed that a request for content from a local user is sent to its local router in the form of an Interest.

\vspace{-0.1in}
\subsection{Information Exchanged and Stored}
\label{sec-info}


The name of content object (CO)  $j$ is denoted by  $n(j)$
and the name prefix that is the best match for  name $n(j)$ is denoted by $n(j)^*$.
The set of neighbors of router $i$ is denoted by $ N^i$.


An Interest forwarded  by router $k$ requesting CO $n(j)$ is denoted by  $I[n(j), AID^I(k), D^I(k) ]$, and states the name  of the requested CO ($n(j)$),  an 
anonymous identifier ($AID^I(k)$) used to denote the 
origin of the Interest, and the distance from $k$ to the requested content.
 
A data packet sent by router $i$ in response to  an Interest
is denoted by  $DP[n(j), AID^R(i), sp(j) ]$, and  states the name  of the  CO being sent ($n(j)$),  an anonymous identifier ($AID^R(i)$) that states the intended recipient of the data packet, and a  security payload ($sp(j)$) used optionally to validate the CO.

A reply sent by router $i$ in response to  an Interest is denoted by $REP[n(j),$   $ AID^R(i)$,  $ \mathsf{CODE} ]$  and states the name  of a CO ($n(j)$),   an anonymous identifier ($AID^R(i)$) that states the intended recipient of the reply, and a code ($\mathsf{CODE}$) indicating the reason why the reply is sent. Possible reasons for sending a reply include: an Interest loop is detected, no route is found towards requested content,  and no content is found.

Router $i$ maintains four tables for forwarding:
an optional  Local Interests Gathered Table ($LIGHT^i$),  
a forwarding information base ($FIB^i$), an anonymous routing table ($ART^i$), and a 
Local Interval Set Table ($LIST^i$).

$LIGHT^i$ lists the names of the COs requested by router $i$
or already stored at router $i$. It is indexed by the CO names that have been requested by the router on behalf of local customers.  
The entry for CO name $n(j)$  states the name of the CO ($n(j)$), 
a pointer to  the content of the CO ($p[n(j)]$), and a list of zero or more identifiers of local consumers ($lc[n(j)]$) that have requested the CO while the content is remote. 

$FIB^i$ is indexed using known content name prefixes.  
The entry for prefix $n(j)^*$ states the  distance reported by each next-hop neighbor router for the prefix. The distance stored for  
neighbor $q$ for prefix $n(j)^*$ in $FIB^i$  is denoted by $D(i, n(j)^*, q)$.
Each entry in $FIB^i$ is stored for a maximum time determined by the lifetime of the corresponding entry in the routing table of the router.


$LIST^i$  maintains the intervals of anonymous identifiers used by router $i$. It states
the local interval of identifiers  accepted by  router $i$ (denoted by $LI^i (i)$), and the  local interval of identifiers 
accepted by  each neighbor router $k$  (denoted by $LI^i (k)$) . 
Clearly,  $LI^i (k) = LI^k (k)$. All local intervals have the same length $|LI|$.

$ART^i$ is indexed using the anonymous identifiers taken from $LI^i (i)$.   
Each entry states an anonymous identifier of a destination
($AID(ART^i)$), 
a next hop to the destination,
$(s(ART^i)$), 
and an identifier mapping used to handle identifier collisions
$(map(ART^i)$).
$ART^i [AID, s, map]$ is used to denote a given  entry in $ART^i$.

Routers can exchange local intervals with their neighbors in a number of ways. The exchange can be done in the data plane using Interests and data packets. An example would be  having a router send an Interest stating a common name denoting that a local interval is requested, and an empty AID. Given the 
succinct way in which local intervals can be stated (an identifier denotes its interval), the exchange can also be easily done  as part of the routing protocol running in the control plane. Routers could exchange interval identifiers in
HELLO messages, link-state advertisements or distance updates. 
To simplify our description of CCN-GRAM,  we  assume that routers have exchanged their local intervals with one another and have populated their LISTs accordingly. 
We also assume  that  local intervals do not change for extended periods of time after they are assigned.

\subsection{Eliminating Forwarding Loops}
\label{lfr}

Let $S^i_{n(j)^*}$ denote the set of next-hop neighbors of router $i$ for prefix $n(j)^*$. The  following  rule  is used to 
ensure that Interests cannot traverse routing loops, even if the 
routing data stored in FIBs regarding name prefixes is inconsistent and leads to routing-table loops.

 \vspace{0.05in}
\noindent
{\bf  Loop-Free Forwarding Rule (LFR):}  \\
Router $i$  
accepts $I[n(j), AID^I(k), D^I(k) ]$ from router $k$ if:
\begin{equation}
\label{lfr}
\exists ~v \in  S^i_{n(j)^*} (~ D^I(k)  > D(i, n(j)^*, v) ~) 
\end{equation}

LFR is based on the same invariants we have proposed previously to
eliminate Interest looping in NDN and CCNx 
\cite{ifip2015, nof2015} and avoid forwarding loops in 
more efficient forwarding planes for content-centric networks
\cite{ocean15}. As we explain in \cite{ifip2015, ocean15, nof2015},
the approach is a simple application of diffusing computations
that  ensures loop-free forwarding of Interests with or without aggregation.

\subsection{Forwarding to  Anonymous Destinations}
 \label{labeling}

The header of a datagram needs to denote its origin and destination, so that the datagram can be forwarded to the intended destination and responses to the datagram can be forwarded back to the source.  Since the introduction of datagram packet switching by Baran \cite{baran}, the identifiers used to denote the sources and destinations of datagrams have  had global scope, and routers maintain FIBs with entries towards those  sources.
However, this need not be the case!  

It is trivial to add information in Interests about the paths they traverse  (e.g., see \cite{source}) to allow responses to be sent back without the need for FIBs maintaining routes to the sources of Interests. However, this would negate the anonymity of Interests advocated in NDN and CCNx.

CCN-GRAM uses local identifiers to denote the sources of Interests in a way that 
responses  to Interests (data packets or replies) can be forwarded correctly to the sources of Interests, without their identity being revealed to relaying routers, caching sites, or content producers.



Given that all local intervals 
have the same length $|LI|$, the local interval $LI^i (i)$ is uniquely defined by the smallest identifier of the interval, which we  denote by $LI^i (i)[s]$.

If router $p$ sends  Interest $I[n(j),  AID^I(p), D^I(p)  ]$ to router $i$, $AID^I(p)$ must be in $LI^p (i) = LI^i (i)$. Similarly, if router $i$ forwards Interest $I[n(j),  AID^I(i), D^I(i)  ]$ to router $n$, $AID^I(i)$ must be in $LI^i(n) = LI^n(n) $. Hence, to forward an Interest from $p$ to $n$, router $i$ must map the AID received in the Interest from $p$ to an AID that belongs to the local interval accepted by its neighbor $n$.
Router $i$ can accomplish this mapping with the following bijection, where $\epsilon$ is a constant known only to router $i$:

\vspace{-0.15in}
{\small
\begin{equation}
\label{bijection}
AID^I(i)  =    \epsilon + AID^I(p) - LI^i(i)[s] + LI^i(n)[s]  ~mod~ |LI|
\end{equation}
}

\vspace{-0.15in}
We denote the mapping of identifiers from $LI^i (i)$ to  $LI^i (n)$ by 
$f_i(n): LI^i (i) \rightarrow LI^i (n).$ The image of identifier $a \in LI^i (i)$ under 
$f_i(n)$ is denoted by $f_i(n)[a]$ and $f_i(n)[a] \in LI^i (n)$.
The reverse mapping from $LI^i (n)$ to $LI^i (i)$ is denoted by $f_i^{-1}(n)$ and 
of course $f_i^{-1}(n)[f_i(n)[a] ] = a$.

Algorithms~\ref{algo-CCN-GRAM-create-Interest}  to \ref{algo-CCN-GRAM-reply} specify 
the steps  taken by  routers to process and forward Interests, and return data packets   or  replies.   
We assume that each router is initialized properly,  knows the identifiers used to denote local consumers, knows all its neighbors, and knows the local identifier intervals associated with each neighbor. 
We assume that a routing protocol (e.g., DCR \cite{dcr}, NLSR \cite{nlsr}) operating in the control plane updates
the entries of  routing tables listing one or multiple next hops towards name prefixes.
Routers populate their FIBs with routes to name prefixes
based on  the data stored in their routing tables. 
How long FIB entries are maintained is determined by the operation of the routing protocol. 

We assume that router $i$ uses a single anonymous identifier in $LI^i(i)$ to denote itself in its $ART^i$, and denote it by $AID^i$. 

\begin{algorithm}[h]
\caption{Processing Interest  from user $c$ at router $i$}
\label{algo-CCN-GRAM-create-Interest}
{\fontsize{7}{7}\selectfont
\begin{algorithmic}
\STATE{{\bf function}  Interest\_Source}
\STATE {\textbf{INPUT:}  $LIGHT^i$, $LIST^i$,  $FIB^i$, $ART^i$,  $AID^i$, $I[n(j), c, nil]$;}

\IF{$n(j) \in LIGHT^i$ }
	\IF{$p[n(j)] \not= nil$ ~(\% CO is local) }
		\STATE{
		retrieve CO $n(j)$; 
		send  $DP[n(j),  c, sp(j) ]$ to consumer $c$}
	\ELSE
		\STATE{$p[n(j)] = nil$; $lc[n(j)]  = lc[n(j)]  \cup c$ ~(\% Interest is aggregated)}
	\ENDIF
\ELSE
	\IF{$n(j)^* \in LIGHT^i$ (\%All content in $n(j)^*$ is local and $n(j)$ is not)}
		\STATE{send  $REP[n(j),  c,  \mathsf{no~ content} ]$ ~(\% $n(j)$ does not exist) }
	\ELSE
		\IF{$n(j)^* \not\in FIB^i$ }
			\STATE{
			send $REP[n(j),  \mathsf{no~ route}, c ]$ to $c$ ~(\% No route to $n(j)^*$ exists)} 
		\ELSE
			\STATE{create  entry for $n(j)$ in $LIGHT^i$: ~(\% Interest from $c$ is recorded) \\
		$lc[n(j)] =  lc[n(j)] \cup c$;  $p[n(j)] = nil$; }

			\IF{$AID^i = nil$} 
				\STATE{
					select identifier   $a \in LI^i(i)$ that is not used in any entry in $ART^i$; }
				\STATE{
										$AID^i = a$; 
					create entry $ART^i[AID^i, i, AID^i]$}
				\ENDIF
			\FOR{{\bf each} $v \in N^i$ {\bf by rank in} $FIB^i$} 
				\STATE{
				$AID^I(i) = f_i(v)[AID^i]$; $D^I(i) = D(i, n(j)^*, v)$; \\
						send $I[n(j), AID^I(i), D^I(i) ]$ to  $v$;  {\bf return}	}
			\ENDFOR	
		\ENDIF
	\ENDIF
\ENDIF
\end{algorithmic}
}
\end{algorithm}

Algorithm \ref{algo-CCN-GRAM-create-Interest}  shows the steps taken by router $i$ to process Interests received from local consumers. For convenience, content requests from local consumers are assumed to be Interests stating  the name of a CO,   the name of the consumer, and an empty distance to the content assumed to denote infinite. Similarly the same format of data packets and replies used among routers is used to denote the responses a router sends to local consumers.

After receiving an Interest from a local consumer, router $i$ first searches its LIGHT to determine if the content is stored locally or a request for the same content is pending. If the content is stored locally, a data packet is sent back to the user requesting the CO.  If a request for the same content is pending, the name of the user is simply added to the list of users that have requested the CO. 

In our description of CCN-GRAM, a router that advertises being an origin  of a prefix must have all the COs associated with the prefix stored locally. 
If router $i$ states that it is an origin of the name prefix $n(j)^*$ and a specific CO with a name that is in that prefix is not found locally, a reply must be sent back to the consumer stating that the content does not exist.  Additional steps could be taken to address the case of Interests sent maliciously for content that does not exist.

If the CO is remote and no FIB entry exists for a name prefix that can match $n(j)$, a reply is sent back stating that no route to the CO could be found. Otherwise, router $i$ forwards the Interest through the highest ranked neighbor $v$ in its FIB for the name prefix matching  $n(j)$, which is denoted by $n(j)^*$.   How such a ranking is done is left unspecified, and can be based on a distributed or local algorithm \cite{dcr, nlsr, gold}.

When router $i$ originates an Interest on behalf of a local consumer and forwards Interest $I[n(j),  AID^I(i), D^I(i)  ]$ to neighbor router $n$ towards name prefix $n(j)^*$,  router $i$ selects an identifier $a \in LI^i (i)$ that is not used to denote any other source of Interests in $ART^i$, sets  $AID^I(i) = f_i(n)[a] \in LI^i (n)$, and stores the entry $ART^i [a, i, a]$.
Router $i$ can use the same anonymous identifier  for all the Interests it originates on behalf of local consumers and forwards to neighbor $n$.

If no ART entry exists with router $i$ as the origin of Interests ($AID^i = nil$), 
$AID^i$ is selected from the set of AIDs in $LI^i(i)$ that are not being used for
other Interest sources, and a new ART entry is created for $AID^i$.
The Interest is forwarded to the selected next hop for the Interest by first
mapping $AID^i$ into an AID in $LI^i(v)$ using  the bijection in Eq. \ref{bijection}.

\begin{algorithm}[h]
\caption{Processing Interest  from router $p$ at router $i$}
\label{algo-CCN-GRAM-Interest}
 {\fontsize{7}{7}\selectfont
\begin{algorithmic}
\STATE{{\bf function} Interest\_Forwarding}
\STATE {\textbf{INPUT:}  $LIGHT^i$, $LIST^i$, $FIB^i$, $ART^i$,  
$I[n(j),  AID^I(p), D^I(p)]$;}
\STATE{$AID^R(i) = AID^I(p)$;}
\IF{$n(j) \in LIGHT^i$ }
	\IF{$p[n(j)] \not= nil$ }
		\STATE{
		retrieve CO $n(j)$;   
		send  $DP[n(j), AID^R(i), sp(j) ]$ to $p$}
	\ENDIF
\ELSE
	\IF{$n(j)^* \in LIGHT^i$}
		\STATE{send  $REP[n(j),  AID^R(i), \mathsf{no~ content} ]$ to $p$ 
		 ~~~(\% $n(j)$ does not exist) }
	\ELSE
		\IF{$n(j)^* \not\in FIB^i$ }
			\STATE{
			send $REP[n(j),  AID^R(i), \mathsf{no~ route} ]$ to $p$ ~(\% No route to $n(j)^*$ exists)} 
		\ELSE

		\FOR{{\bf each} $s \in N^i$ {\bf by rank in} $FIB^i$} 
			\IF {$ D^I(p)  > D(i, n(j)^*, s) $   ~(\% LFR is satisfied) }
				\STATE{$SET = \emptyset$;  $AID^I(i) = nil$; $collision = 0$;}
				\FOR{{\bf each} entry $ART^i[AID, s, map]$} 
					\STATE{$SET  = SET \cup \{ AID \}$;}
					\IF{$AID(ART^i ) = AID^I(p)$}
						\IF{$s(ART^i) = p$}
							\STATE{$AID^I(i) = f_i(s)[AID(ART^i) ]$}
						\ELSE
						\STATE{$collision = 1$}
						\ENDIF
					\ENDIF
					\IF{$map(ART^i ) = AID^I(p) \wedge s(ART^i) = p$}
						\STATE{$AID^I(i) = f_i(s)[AID(ART^i) ]$}
					\ENDIF
				\ENDFOR
				\IF{$collision = 0 \wedge AID^I(i) = nil$}
					\STATE{
					create entry $ART^i[AID^I(p), p, AID^I(p)];$}
					\STATE{$AID^I(i) = f_i(s)[AID^I(p)]$}
				\ENDIF
				\IF{$collision = 1 \wedge AID^I(i) = nil$}
					\STATE{
					select $a \in LI^i(i) - SET$;  \\
					create entry $ART^i[a, p, AID^I(p)];$
					$AID^I(i) = f_i(v)[a]$}
				\ENDIF
				\STATE{
				$D^I(i) = D(i, n(j)^*, s) $; \\
				send $I[n(j), AID^I(i), D^I(i)]$ to  $s$; {\bf return}}
			\ENDIF
		\ENDFOR
			 ~(\% LFR is not satisfied; Interest may be traversing a loop)
				\STATE{
				send $REP[n(j),  AID^R(i), \mathsf{loop} ]$ to $p$} 
		\ENDIF
	\ENDIF
\ENDIF

\end{algorithmic}
}
\end{algorithm}

Algorithm~\ref{algo-CCN-GRAM-Interest} shows the steps taken by router $i$ to process an Interest received from a neighbor router $p$. The main differences in the steps taken by router $i$ compared to Interests received from local users are that no Interest aggregation is done for Interests received from neighbor routers, and router $i$ maps the AID it receives in the Interest from the previous hop to the AID it should use in the Interest it sends to the next hop using a simple mapping function.

When router $i$ forwards  Interest $I[n(j),  AID^I(p), D^I(p) ]$ from predecessor router $p$ to successor router $n$ towards name prefix $n(j)^*$, router $i$ 
makes sure that $AID^I(p) \in LI^i (i)$ is not listed in an $ART^i$ entry with a next hop other than $p$. If that is the case, 
router $i$ stores 
$ART^i [AID^I(p), p, AID^I(p)]$, and sets 
$AID^I(i) = f_i(n)[AID^I(p)] \in LI^i (n)$. Otherwise,
router selects an AID  $ b \in LI^i (n)$ that is not used 
to denote any other source of Interests in $ART^i$, 
stores $ART^i [b, p, AID^I(p)]$, and sets 
$AID^I(i) = f_i(n)[b] \in LI^i (n)$. 

If the requested content is cached locally, a data packet  is sent back. If router $i$ is  an origin of $n(j)^*$ and the CO with name $n(j)$ is not found locally,  a reply is sent back stating that the content could not be found. Additional steps can be taken to address the case of malicious Interests requesting non-existing content.
If the CO is remote and no FIB entry exists for $n(j)^*$, then router sends a reply stating that no route could be found for the CO. 

Router $i$ tries to forward the Interest to a next hop $s$ for the best prefix match for $n(j)$ that satisfies LFR. 
The highest-ranked router 
satisfying LFR is selected as the successor for the Interest and router $i$.
If no neighbor is found that satisfies LFR, a reply is sent stating that a loop was found.

\begin{algorithm}[h]
\caption{Processing data packet from router $s$ at router $i$}
\label{algo-CCN-GRAM-Data}
{\fontsize{7}{7}\selectfont
\begin{algorithmic}
\STATE{{\bf function} Data Packet}
\STATE{\textbf{INPUT:}  $LIGHT^i$,  $LIST^i$, $ART^i$, 
$DP[n(j),  AID^R(s), sp(j) ]$; }
\STATE{{\bf [o]} verify $ sp(j)$;}
\STATE{{\bf [o]} {\bf if} verification with $ sp(j)$ fails {\bf then} discard $DP[n(j), AID^R(s), sp(j) ]$;}

\STATE{$a = f_i^{-1}(s)[  AID^R(s)  ] $;  retrieve entry $ART^i[a, p, m ]$;
}
\STATE{{\bf if} $ART^i[a, p, m ]$ does not exist {\bf then} drop $DP[n(j), AID^R(s), sp(j) ]$;}
\IF{ $p = i$~~(\% router $i$ was the origin of the Interest) }
	\FOR{{\bf each} $c \in lc[n(j)]$}
		\STATE{send $DP[n(j),c,  sp(j) ]$ to  $c$; $lc[n(j)] = lc[n(j)] - \{ c \}$}
	\ENDFOR
\ELSE
	\IF{$p \in N^i $}
		\STATE{
			$AID^R(i)= m$; send $DP[n(j), AID^R(i), sp(j)  ]$ to  $p$}
	\ENDIF
\ENDIF

\STATE{
		{\bf if} no entry for  $n(j)$ exists in $LIGHT^i$ {\bf then} \\
		~~~create  $LIGHT^i$ entry for $n(j)$: $lc[n(j)] = \emptyset$  \\ {\bf end if}\\
	store CO in local storage;   $p[n(j)] =$ address of CO in local storage
	} 

\end{algorithmic}}       
\end{algorithm}

Algorithm~\ref{algo-CCN-GRAM-Data} outlines the processing of data packets.  
If  local consumers requested the content in the data packet, it is sent to those consumers based on the information stored in $LIGHT^i$. If the data packet is received in response to an Interest that was forwarded from router $p$, router $i$ forwards the data packet  doing the proper mapping of AIDs.  Router $i$ stores the data object if edge or on-path caching is supported. 

When router $i$ receives $DP[n(j), AID^R(n), sp(j) ]$ from neighbor $n$, it 
obtains the AID of of the destination where the packet should be forwarded by computing  $f_i^{-1}(n)[AID^R(n)]$.   Router $i$ uses entry  $ART^i[  f_i^{-1}(n)[AID^R(n)], p, m]$ to determine the next-hop neighbor $p$ that should receive the data packet, and sets $AID^R(i) = m$.

\begin{algorithm}[h]
\caption{Process reply from router $s$ at router $i$}
\label{algo-CCN-GRAM-reply}
{\fontsize{7}{7}\selectfont
\begin{algorithmic}
\STATE{{\bf function} REPLY}
\STATE {\textbf{INPUT:}  $LIGHT^i$,  $LIST^i$, $ART^i$, 
$REP[n(j), AID^R(s), \mathsf{CODE}]$; 
}

\STATE{$a = f_i^{-1}(s)[  AID^R(s)  ] $;  retrieve entry $ART^i[a, p, m ]$;
}
\STATE{{\bf if} $ART^i[a, p, m ]$ does not exist {\bf then} drop $REP[n(j), AID^R(s), \mathsf{CODE} ]$;}
\IF{ $p = i$~~(\% router $i$ was the origin of the Interest) }
	\FOR{{\bf each} $c \in lc[n(j)]$}
		\STATE{send $REP[n(j), c, \mathsf{CODE}]$ to  $c$}
	\ENDFOR
	\STATE{delete entry for $n(j)$ in $LIGHT^i$}
\ELSE
	\IF{$p \in N^i $}
		\STATE{
			$AID^R(i)= m$; send $REP[n(j), AID^R(i),  \mathsf{CODE} ]$ to  $p$}
	\ENDIF
\ENDIF

\end{algorithmic}}
\end{algorithm}

 Algorithm~\ref{algo-CCN-GRAM-reply} states the steps taken to handle replies, which are similar to the forwarding steps taken after receiving a data packet. 
Router $i$ forwards the reply to local consumers  if it was the origin of the Interest, or  to a neighbor router $p$  if it has an ART entry  with $p$ as the next hop towards the destination denoted by the AID stated in the reply.

 \subsection{Example}

\begin{figure}[h]
\begin{centering}
    \mbox{
    \subfigure{\scalebox{.2}{\includegraphics{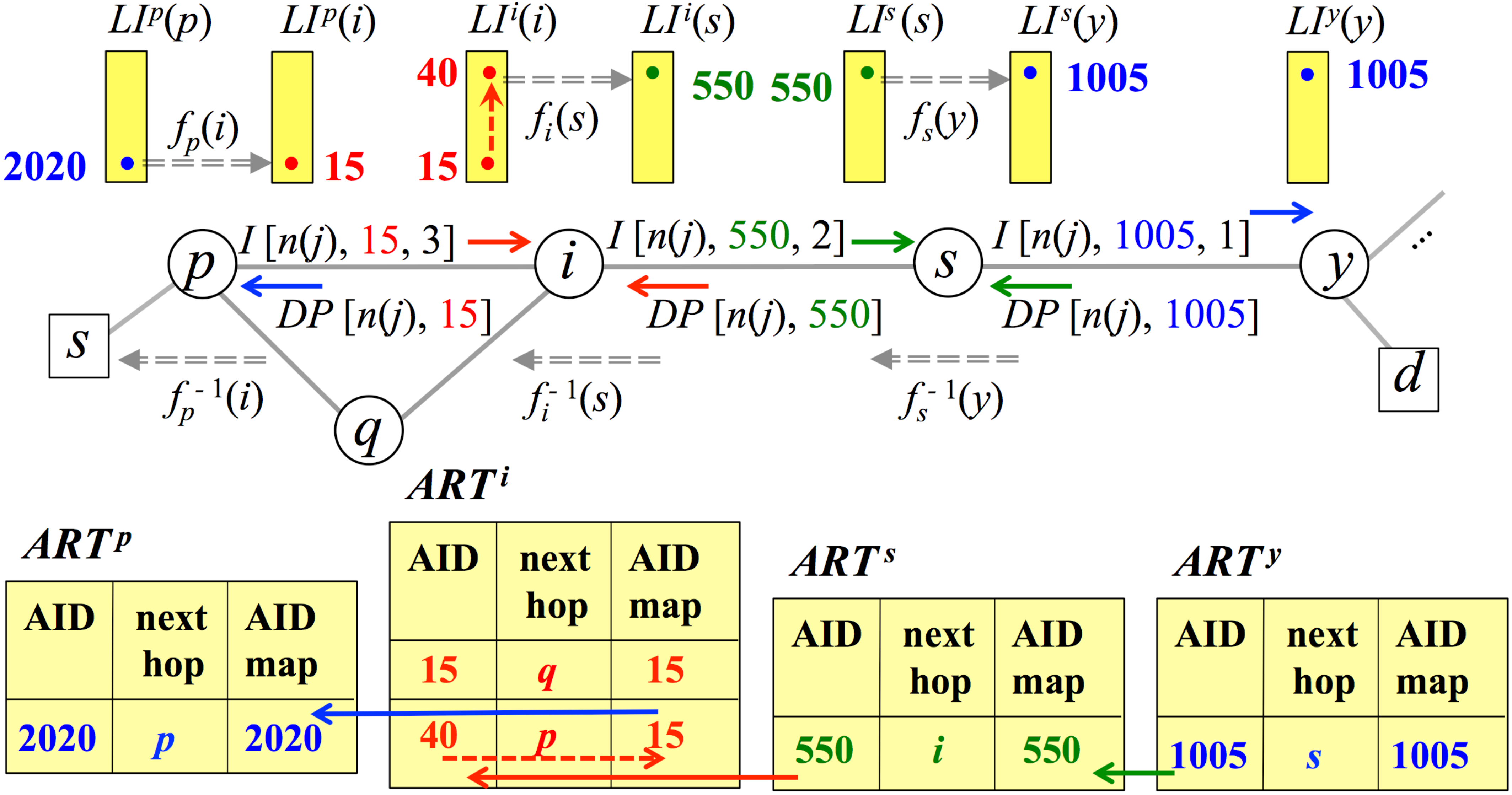}}}
      }
\vspace{-0.2in}
   \caption{Forwarding of Interests and responses to them in CCN-GRAM
   }
   \label{gram1}
\end{centering} 
\end{figure}

\vspace{-0.05in}
Figure~\ref{gram1} illustrates the swapping of AIDs used by routers to forward Interests and responses to them. The local intervals used in the figure are small
for simplicity, and the figure focuses on the forwarding state needed to forward Interests from $p$ to name prefixes announced by router $y$,  as well as the responses to such Interests. Interests are forwarded based on FIB entries, and 
responses to Interests (data packets or replies)  are forwarded based on ART entries.

As illustrated in Figure \ref{gram1}, router $i$ takes into account the possibility of collisions in the AIDs stated in Interests received from  different neighbors by means of the identifier-mapping filed of ART entries.
The bijection in Eq. \ref{bijection} is used to map either the AID specified in the Interest received from neighbor $p$ or the AID created by router $i$ to handle collisions 
to the AID stated by router $i$ in the Interest it forwards to a next-hop router $s$.
In the example, router $i$ has an exiting entry $ART^i [15, q, 15]$  when it receives  Interest $I[n(j), 15, 3 ]$ from router $p \not= q$. Accordingly, router $i$ selects $AID = 40$, creates  entry 
$ART^i[ 40, p, 15]$, and sets  $AID^I(i) = f_i(s)[40] = 550$ before forwarding
Interest $I[n(j), 550, 2 ]$ to router $s$.  When router $i$ receives data packet 
$DP[n(j), 550, sp(j) ]$ from router $s$, it computes $f_i^{-1}(s)[550] = 40$.
Using $AID = 40$ as the key in $ART^i$, router $i$ obtains the next hop $p$,
sets $AID^R(i) = 15$, and forwards 
$DP[n(j), 15, sp(j) ]$ to router $p$.

It is clear from the example that a router sending an Interest is unaware of collisions of AIDs at the next hop. The identifier mapping field of ARTs 
allows routers to multiplex Interests from different neighbors stating the same AID values.

Even when a very small number of routers is involved, only the router that originates an Interest is able to determine that fact, because the identifiers used for Interest forwarding are assigned by the next hops.




\subsection{Native Support for Multicasting}

Support of multicast communication in the data plane with no additional signaling required in the control plane is viewed as an
important benefit derived from maintaining per-Interest forwarding state using PITs. In short, multicast receivers send Interests towards the multicast source.
As Interests from receivers are aggregated in the PITs  on their way to the multicast source, a multicast  forwarding tree (MFT)  is formed and maintained in the data plane. Multicast Interest are forwarded using the same FIB entries used for unicast traffic,
and  multicast data packets are sent using  reverse path forwarding (RPF)  over the paths traversed by aggregated Interests. 
Using PITs is appealing in this context; however,   as we show below, native  support of multicasting  in the data plane
can be easily done with no need for per-Interest forwarding state! 


\subsubsection{Information Stored and Exchanged}


We assume that the name stated in an Interest created to request content from a multicast source  denotes 
a multicast source uniquely, and call such an Interest a {\em multicast Interest}. We also assume that consumers and routers
differentiate between a multicast Interest and  an Interest  originated from a single consumer (unicast Interest).


A multicast Interest $MI[g(j),  D^I(i), mc^I(i)]$ sent by router $i$ to router $n$ 
states: the name of a multicast group $g(j)$, the distance from router $i$  to the source of the multicast group $D^I(i)$, and a multicast counter ($mc^I(i)$) used for pacing. 

A multicast data packet   $MP[g(j),  sp(j), mc^R(i)]$ states the name of the multicast group $g(j)$,  a security payload $sp(j)$, a multicast counter $mc^R(i)$, 
plus the content payload. 
A multicast reply
$MR[g(j), CODE, mc^R(i)]$  states the reason for the reply and the current value of the multicast counter.

Router $i$ maintains a  multicast anonymous routing table ($MART^i$) that contains the  forwarding state to the receivers of  multicast groups. Each entry in $MART^i$ specifies a multicast group name,  the value of the multicast counter ($mc$), and a list of next hops to the group of receivers who have sent Interests for the group.  If router $i$  has local receivers for group $g(j)$, the entry for the group in $MART^i$  includes router $i$ as a next hop to the receivers of the group. 

Router $i$ also maintains a group membership table ($GMT^i$) that lists the mappings of multicast group names to the lists of local receivers that requested to join the groups. The GMT entries allow the router to deliver 
multicast content to local receivers of specific groups.

\subsubsection{Multicast Content Dissemination}

\begin{figure}[h]
\begin{centering}
    \mbox{
    \subfigure{\scalebox{.22}{\includegraphics{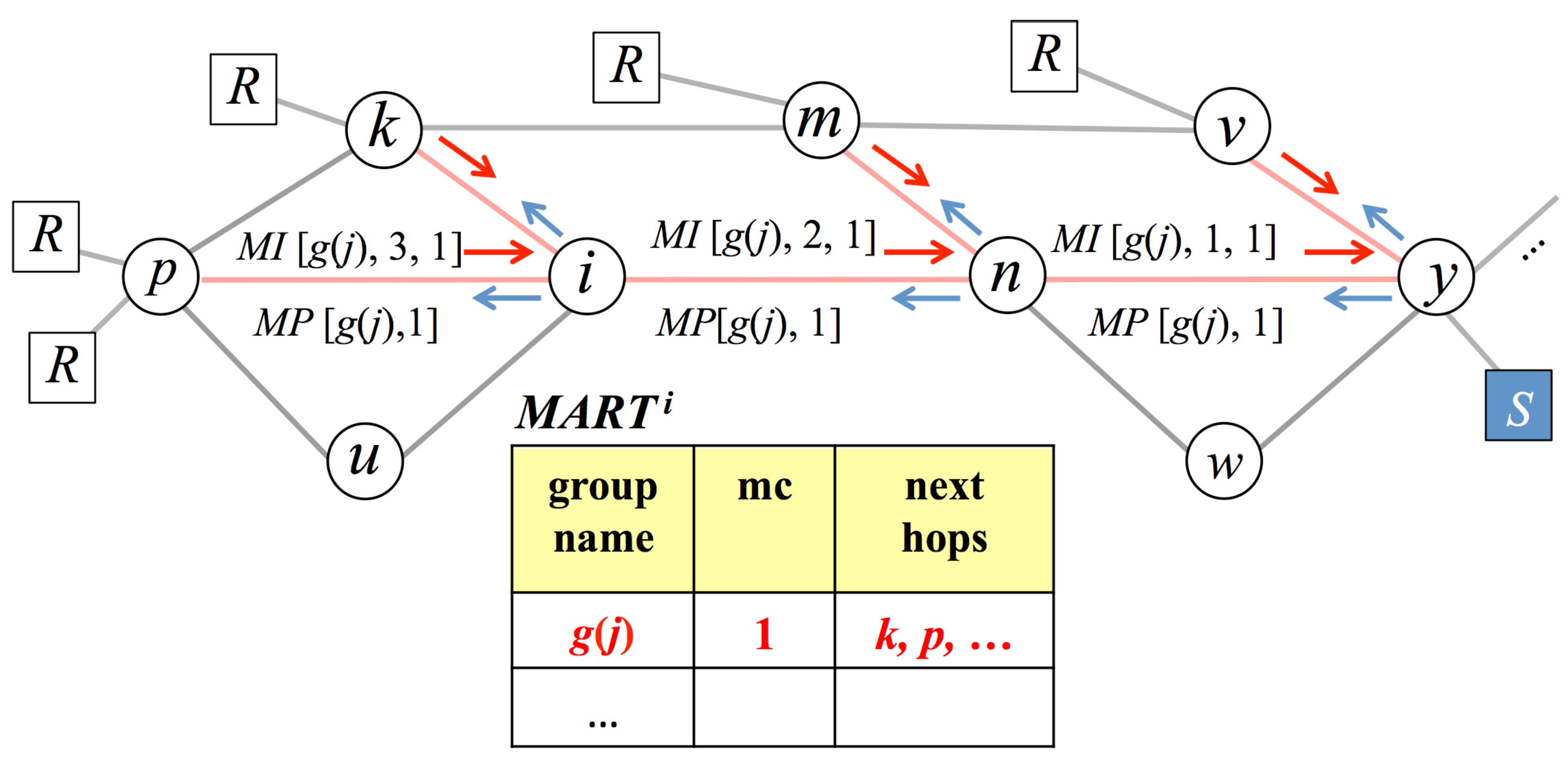}}}
      }
\vspace{-0.08in}
   \caption{Native multicast support in CCN-GRAM
   }
   \label{gram3}
\end{centering} 
\end{figure}

The  key difference of the way in which CCN-GRAM forwards multicast traffic compared to NDN or CCNx is that a MART maintains per-group forwarding state, while a PIT maintains per-Interest forwarding state. Figure \ref{gram3} illustrates the forwarding of multicast  Interests and multicast content  in CCN-GRAM. There is no need for anonymous identifiers for multicast  content forwarding, because all consumers of a group must receive the same multicast COs, which are forwarded using multicast group names.



A content consumer $c$ requests  to join a multicast group $g(j)$ as a receiver by sending a multicast Interest $MI[g(j), D^I(c), mc^I(c)]$ with  $D^I(c) = nil$.

If router $i$ has multiple  local receivers or neighbor routers requesting to join the same multicast group  $g(j)$, router $i$ forwards multicast Interest $MI[g(j), D^I(i), mc^I(i)]$  only once  towards the source of the multicast group $g(j)$ based on the information in its FIB. 
Router $i$ simply adds new local consumers to the entry for $g(j)$ in  $GMT^i$ or new next hops to multicast receivers in $MART^i$.  

Router $i$ forwards  multicast data packets based on the group names stated in the packets and the next hop stored in its MART entries, and discards the data packet  if no MART entry exists for the multicast group. A similar approach is used for replies to Interests regarding multicast groups.

The dissemination of multicast data packets over the MFT of a multicast group can be of two types. A multicast source can push multicast data towards the receivers, or the receivers can pull data from the source by submitting Interests.
  
{\bf Push-based dissemination:} The only forwarding state needed in CCN-GRAM for push-based multicast dissemination consists of the name of a multicast group and the names of the next hops towards the group receivers. In this mode, the $mc$ value of an entry in a MART is updated with each multicast data packet forwarded by the router towards the receivers.

{\bf Pull-based dissemination:} CCN-GRAM can also support pull-based multicast dissemination with no need for  per-Interest forwarding state.
An exemplary approach consists of  a source-pacing algorithm based on the  $mc$ values carried in Interests and data packets.
Each receiver increments the $mc$ value of Interests it sends for the group asking
for the next piece of multicast content from the source.  When router $i$ receives multicast  Interest  $MI[g(j), D^I(p), mc^I(p)]$ from a neighbor router or a local content consumer $p$, it forwards the Interest only if $mc = 1 + v $, where $v$ is the current $mc$ value stored in $MART^i$ for the multicast group. Router $i$ updates the $mc$ value in $MART^i$ as it forwards the Interest, and subsequent Interests with the same $mc$ value of $1 + v $ are simply dropped.  As a result,  each router in an MFT forwards a single copy of any Interest  asking for the next multicast content object towards the source. This is like   aggregating 
Interests for a multicast group over the MFT of the group, but with no need to store per-Interest forwarding state.




\section{Performance Comparison}
\label{sec-perf}

We compare the forwarding entries needed to forward Interests and responses in NDN and CCN-GRAM, as well as the end-to-end delays incurred, using simulation experiments  based on implementations of NDN and CCN-GRAM in the ndnSIM simulation tool \cite{ndnsim}.  The NDN implementation was used without modifications, and CCN-GRAM was implemented in the ndnSIM tool following Algorithms 1 to 4. 

The network topology consists of 150 routers distributed uniformly in a 100m $\times$ 100m area and routers with distance of 15m or less are connected with point-to-point links of delay 15ms. The data rates of the links are set to 1Gbps to eliminate the effects that a sub-optimal implementation of  CCN-GRAM or NDN may have on the results.  Only 10 routers chosen randomly are connected to  local content producers of multiple name prefixes, 50 other routers are connected to local content consumers, and all routers act as relays. This choice was made to illustrate the existence of a ``network edge"  and the fact that only a relatively small number of sites host content producers.   Interests are generated with a Zipf distribution with  parameter $\alpha = 0.7$ and producers are assumed to publish 1,000,000 different COs. Each cache can store up to 1000 objects, or 0.1\% of the content published in the network. This caching capacity was selected to  compare on-path caching with edge caching when Interests must be forwarded in the network, rather than being answered with locally cached content.


We considered {\it total Interest  rates per router} of  50, 100, 500, and 2000 objects per second corresponding to the sum of Interests from the local consumers connected to a router. 
The increasing values of total request rates can be viewed as higher request rates from a constant user population of local active users per router, or an increasing  population of active users per router. 
The Interest rates we assume are actually very low according to recent results addressing the size that PITs would have under realistic Internet settings  \cite{dai-12, var-13, vir-13, tsi14}. 

We considered  on-path caching and edge caching. For the case of on-path caching, every router on the path traversed by a data packet from the producer to the consumer caches the CO in its  local cache. On the other hand, with edge caching, only the router directly connected to the requesting  consumer  caches the resulting CO. All caches are LRU.

 \subsection{Size of Forwarding Tables}

\begin{figure}[h]
\begin{centering}
    \mbox{
    \subfigure{\scalebox{.19}{\includegraphics{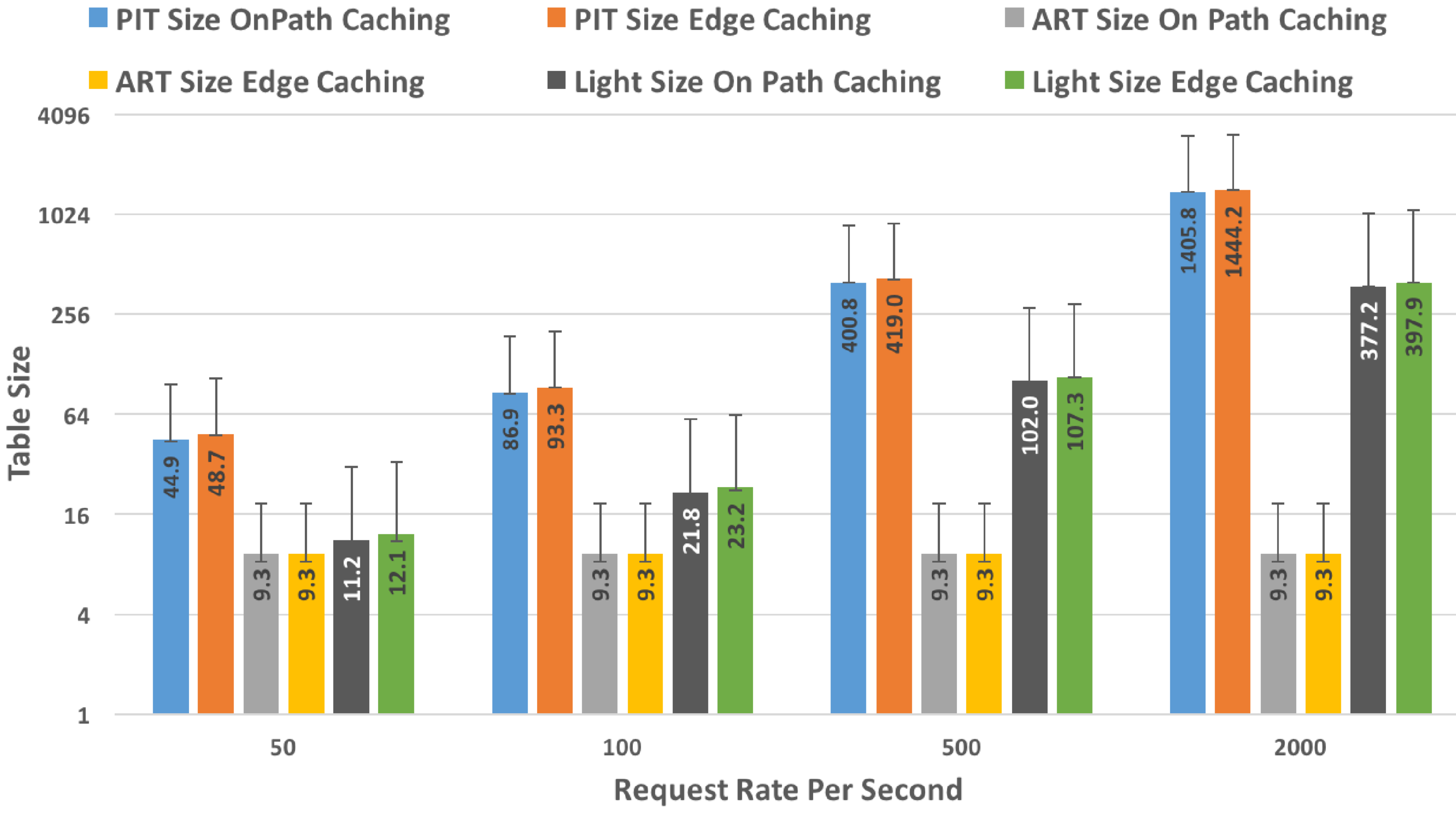}}}
    }
\vspace{-0.2in}
   \caption{Average size of forwarding tables }
   \label{sizePitVSArt}
\end{centering} 
\end{figure}

Figure \ref{sizePitVSArt} shows the average size and standard deviation of the sizes of  PITs,  ARTs and LIGHTs on a logarithmic scale as functions of Interest rates. The size of LIGHTs corresponds only to 
the number of local  Interests pending responses. The number of entries corresponding to content cached locally can be up to 1000 for both NDN and CCN-GRAM.

\vspace{-0.05in}

\vspace{-0.2in}
\begin{figure}[h]
\begin{centering}
    \mbox{
    \subfigure{\scalebox{.19}{\includegraphics{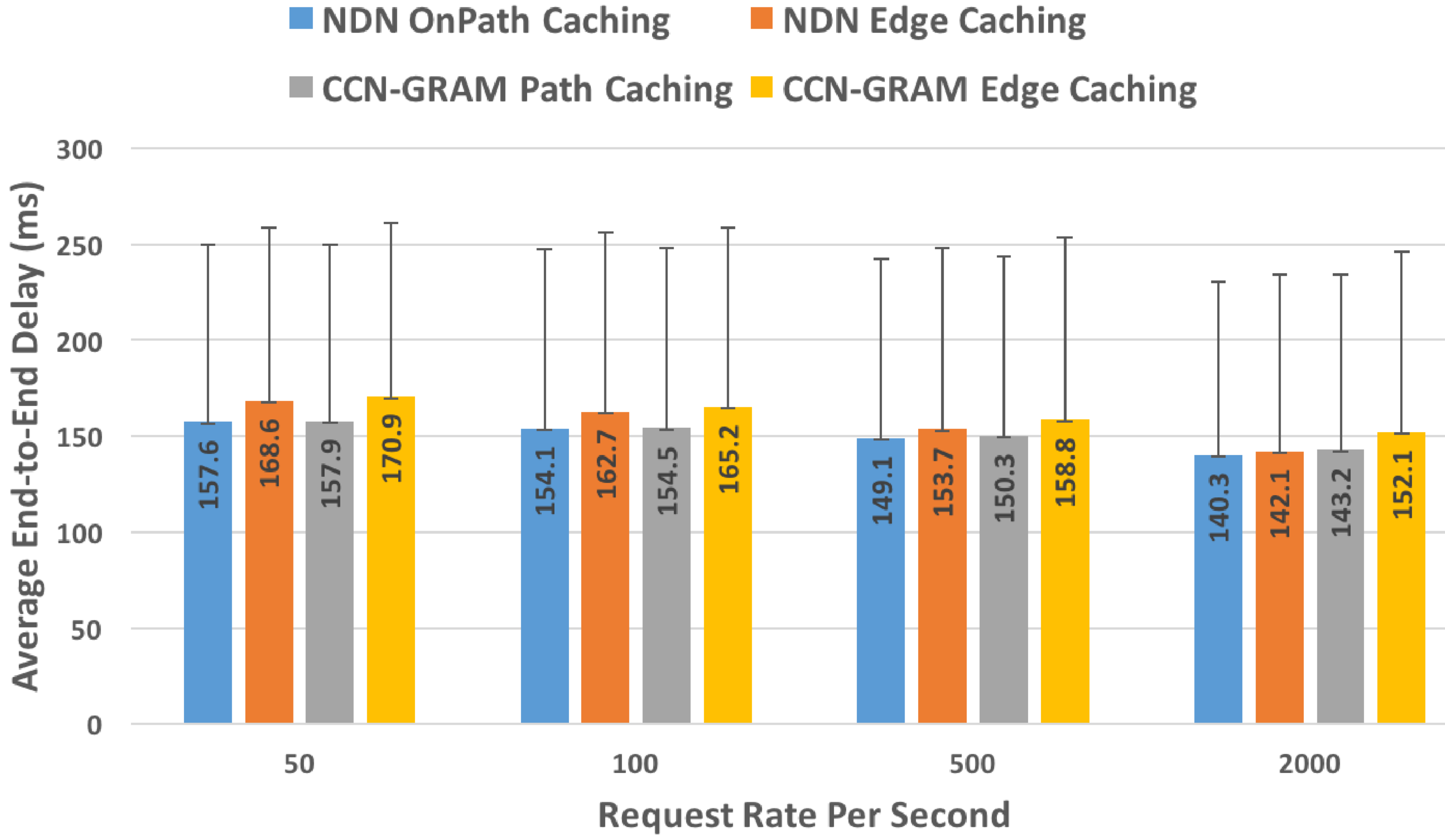}}}
    }
\vspace{-0.25in}
   \caption{Average end-to-end delays}
   \label{delayOCEANvsNDN}
\end{centering} 
\end{figure}

As the figure shows, the size of PITs grows dramatically as the rate of content requests increases, which is expected given that PITs maintain per-Interest forwarding state. By contrast, the size of ARTs, which is the only forwarding state stored  by relay routers,  is only a small fraction of the total number of routers and remains fairly constant with respect to the content request rates, which is always one or multiple orders of magnitude smaller than the average PIT size. The size of LIGHTs is a function of the number of COs requested locally or cached on path, but the average size of a LIGHT is an order of magnitude smaller than the average size of a PIT.
The size of a ART is independent of where content is being cached, given that an ART entry  is stored independently of how many Interests traverse the route.   Interestingly, edge-caching renders only slightly larger PIT sizes than on-path caching in NDN.


\subsection{Average Delays}

Figure \ref{delayOCEANvsNDN} shows the average end-to-end delay for NDN and CCN-GRAM as a function of   content request rates for on-path caching and edge caching.  As the figure shows, the average delays for NDN and CCN-GRAM are  
comparable for all values of the content request rates.  This should be expected, given that the  static, loop-free routes in the FIBs prevent Interests to ``wait to infinity" in PITs,  the signaling overhead incurred  by NDN and CCN-GRAM is similar, and in-network caching obviates the need for Interest aggregation.


\section{Conclusions and Future Work}

We presented simulation results showing that Interest aggregation rarely occurs when in-network caching is used.  
Our analysis is limited; however, our detailed  characterization of  Interest aggregation 
via analytical modeling and simulation analysis \cite{ali-ifip16} renders the same conclusion.

We  introduced CCN-GRAM to eliminate the performance limitations associated with  PITs.
CCN-GRAM is the first approach to Interest-based content-centric networking that supports the forwarding of Interests and responses to them using  datagrams that do not reveal the identity of their origins to forwarding routers, caching sites, or content providers.

Simulation experiments were used to show that end-to-end delays  incurred in  CCN-GRAM and NDN are similar when either edge caching or on-path caching is used, but  the 
storage requirements for  CCN-GRAM are orders of magnitude smaller than for NDN.  The results for CCN-GRAM  indicate that it could be deployed with only routers at the edge maintaining LIGHTs
and caches. 
Additional work is needed to make the forwarding of Interests in CCN-GRAM as efficient as the forwarding of responses to Interests using ARTs. The goal is to enable Interest forwarding at Internet scale that does not require
routers to look up FIBs with billions of name-prefix entries as is the case in NDN and CCNx.


Both ARTs and PITs  must be updated when the paths traversed by Interests and their responses must change due to congestion, topology changes, or mobility of consumers and providers. 
Yi et al \cite{ndn-fw2} argue that per-Interest forwarding state  enables faster response to topology changes and congestion, because local repair 
mechanisms can be used.  However, 
multipath routing, and dynamic load balancing 
schemes based on datagram forwarding have been shown to attain results very close to  optimal routing   \cite{vutukury} and can be easily applied to CCN-GRAM in the future. 


CCN-GRAM  can use the same content security features adopted in CCNx and NDN to limit or eliminate cache poisoning attacks, because it makes no modifications to the way in which content is protected in data packets or how a name can be securely linked to the payload of a CO.  However, CCN-GRAM enjoys an enormous  advantage over CCNx and NDN in that it  eliminates the ability for malicious users to  mount Interest-flooding attacks aimed at overwhelming the forwarding tables of routers  \cite{DDos1, vir-13}.
An ART entry  can be added only for valid local identifiers at each router and for routes that satisfy the ordering constraint imposed with LFR.  Given that both conditions are managed in the control plane,  mounting  attacks on ARTs is 
much more difficult  than simply having users send Interests for COs corresponding to valid name prefixes. 



 {\fontsize{7}{7}\selectfont

  }

\end{document}